\documentclass{eas}
\usepackage{graphicx}
\usepackage{astron}

\begin{document}

\title{NEWAGE} 
\author{K.~Miuchi}\address{Cosmic-Ray Group, Department of Physics, 
Kyoto University 
Kyoto, 606-8502, Japan
\email{miuchi@cr.scphys.kyoto-u.ac.jp}}
\author{K.~Nakamura}\sameaddress{1}
\author{A.~Takada}\sameaddress{1}
\author{S.~Iwaki}\sameaddress{1} 
\author{H.~Kubo}\sameaddress{1}
\author{T.~Mizumoto}\sameaddress{1} 
\author{H.~Nishimura}\sameaddress{1}
\author{J.~Parker}\sameaddress{1}
\author{T.~Sawano}\sameaddress{1} 
\author{T.~Tanimori}\sameaddress{1}
\author{H.~Sekiya}\address{
Kamioka Observatory, ICRR, The University of Tokyo
Gifu, 506-1205 Japan}
\author{A.~Takeda}\sameaddress{2}
\author{T.~Fusayasu}
\address{Department of Human and Computer Intelligence,
Faculty of Informatics, Nagasaki Institute of Applied Science,
Abamachi 536, Nagasaki 851-0193, Japan}
\author{A.~Sugiyama}
\address{Saga University, Saga, Japan }
\author{M.~Tanaka}\address{Institute of Particle and Nuclear Studies, KEK, Tsukuba, Japan}

\begin{abstract}
NEWAGE is a direction-sensitive dark matter search experiment 
with a gaseous time-projection chamber.
We improved the direction-sensitive dark matter limits by our 
underground measurement. 
In this paper, R\&D activities sinse the first underground measurement are described.

\end{abstract}
\maketitle
\section{Introduction}
Direction-sensitive dark mater search with gaseous detector 
was proposed in late 1980s\cite{ref:TPCforDM2,ref:TPCforDM3}.
Since then several experimental and theoretical works on
the possibility of detecting this distinct signal of dark matter 
have been performed\cite{ref:CYGNUS_whitepaper} and the 
references therin. Among these proposed methods, 
The DRIFT group has pioneered studies of gaseous detectors 
for WIMP-wind detection for more than ten years with multi-wire 
proportional chambers \cite{ref:DRIFT_NIM2009}.

We started a new project,  NEw  generation  WIMP-search
with Advanced Gaseous tracking device Experiment (NEWAGE) 
in 2003\cite{ref:NEWAGE_PLB2004}. 
We adopted a new technology named Micro-Patterned Gaseous Detector (MPGD) 
and thus had advantages in the pitch of the detection sensors and a 
three-dimensional tracking scheme. 
We performed first direction-sensitive dark matter search experiment in 
a surface laboratory \cite{ref:NEWAGE_PLB2007} and 
updated the direction-sensitive limits by the measurement 
in an underground laboratory\cite{ref:NEWAGE_PLB2010}.
Although we set a WIMP-proton cross section of 5400 pb for 150 GeV WIMPs by 
a direction-sensitive methods, we need to improve more than three orders 
of magnitude to set a competitive limits to other direction-insensitive 
searches.
We plan to improve the sensitivity by 
reducing the internal radioactive background, 
lowering the energy threshold,  
discriminating head-tails of nuclear tracks, building large-volume detectors, 
and developing the pixel readout.
In this paper, we report these efforts 
after our first underground run.

\section{NEWAGE detectors}
We have three time projection chambers (TPCs) and two Radon Detectors.
We list the specifications of our TPCs in Table~\ref{tab:NEWAGE_TPCs}.
We have one large TPC in the underground laboratory for dark matter run
and background studies.
We have one large and one small TPC in the surface laboratory for 
mainly large volume R\&D and advanced tests, respectively.
We use $\mu$-PIC\cite{ref:10uPIC,ref:30uPIC}, one of the micro 
patterned gaseous detector, as a main-multiplier and a readout of the TPC.
The pitch of the $\mu$-PIC is 400 $\mu$m.
We also use gas an electron multiplier (GEM)\cite{ref:GEM} as a 
sub-multiplier. 
For details on our TPC system, please refer to our previous 
publication\cite{ref:NEWAGE_PLB2010} and the references therein.
We developed two electrostatic collection radon detectors 
similar to the system developed by the 
Super-Kamiokande group\cite{ref:SK_Radon}.
Our detectors are smaller in size 
( $\rm diam. 22.5cm \times 15cm $ and  $\rm diam. 22.3cm \times 15cm $ )
We operate the radon detector with the $\rm CF_4$ at 152 Torr
to measure the radon emanation into the TPC gas.

\begin{table}
\begin{tabular}{|l|l|l|l|l|}
\hline
name & $\mu$-PIC& drift  & location & use\\
\hline
NEWAGE-0.1a& $\rm 10 \times 10 cm^2$& 10cm& surface &advanced test\\
\hline
NEWAGE-0.3a& $\rm 30 \times 30 cm^2$& 30cm&underground &DM run, BG study \\
\hline
NEWAGE-0.3b& $\rm 30 \times 30 cm^2$& 50cm&surface & large volume \\ 
\hline
\end{tabular}
\caption{\label{tab:NEWAGE_TPCs}
Specifications of NEWAGE TPCs.
}
\end{table}




\section{Underground R\&D}
One of the major background in our detector is radons 
coming from uranium and thorium contaminations in 
the detector materials. 
We attached a mini-chamber containing 
about 100 gram of charcoal (TSURUMICOAL 2GS) 
as radon filter 
and circulated the TPC gas with a 
Teflon bellows pump (TSURUMICOAL 2GS). 
With this radon elimination system, 
radon rate was decreased to less than 1/10. 
 
We measured the radon emanation from main components of 
our detector with our radon detectors.
As the absolute detection efficiency of the radon-daughter ion 
was not measured yet, we relatively compared the contribution of each 
components.
Measured results normalized to the NEWAGE-0.3a detector is shown in 
Figure \ref{tab:RD_results}.
We found the glass-reignforced fluoroplastic (TPC cage) had the largest 
radon emanation. 
We decided to replace it with PTFE.
The picture of our new TPC cage is shown in Fig.~{fig:PTFE-TPC}.
The radon background is now expect to be less than 1/3. 
With these major and minor improvements, we started a dark matter run 
on August 3rd, 2011.

\begin{table}
\begin{tabular}{|l|l|}
\hline
material & radon rate per NEWAGE-0.3a (a.u.)\\
\hline
glass-reignforced fluoroplastic (TPC cage) & 1 \\ \hline
PTFE (TPC cage) & 0.1$>$ \\ \hline
glass-reignforced plastic (GEM frame) & 0.7 $\times 10^{-2} >$ \\ \hline
polyimide+copper (GEM) & 0.1$>$ \\ \hline
resistors (TPC) & 0.8 $\times 10^{-2} >$ \\ \hline
$\mu$-PIC &  $0.1 >$ \\ \hline

\end{tabular}
\caption{\label{tab:RD_results}
Relative contribution of the detector components to the radon 
background.
}
\end{table}

\begin{figure}
\includegraphics[width=.5\linewidth]{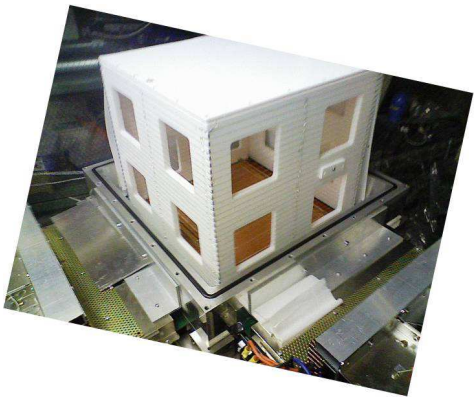}
\caption{\label{fig:PTFE-TPC}
Picture of NEWAGE-0.3a detector with a new drift cage made of PTFE.
}
\end{figure}

\section{Surface R \&D (head-tail recognition)}

Head-tail recognition of the nuclear track is 
important to improve the sensitivity 
of a direction-sensitive dark matter search experiment\cite{ref:Anne}.
DM-TPC group has shown the possibility of 
head-tail recognition of high energy nuclear 
tracks\cite{ref:DMTPC_NIM2008} 
followed by the DRIFT group's work in the 
energy range relevant to the dark matter search for 
one dimension\cite{ref:DRIFT_headtail_exp}.

\begin{figure}
\includegraphics[width=1.\linewidth]{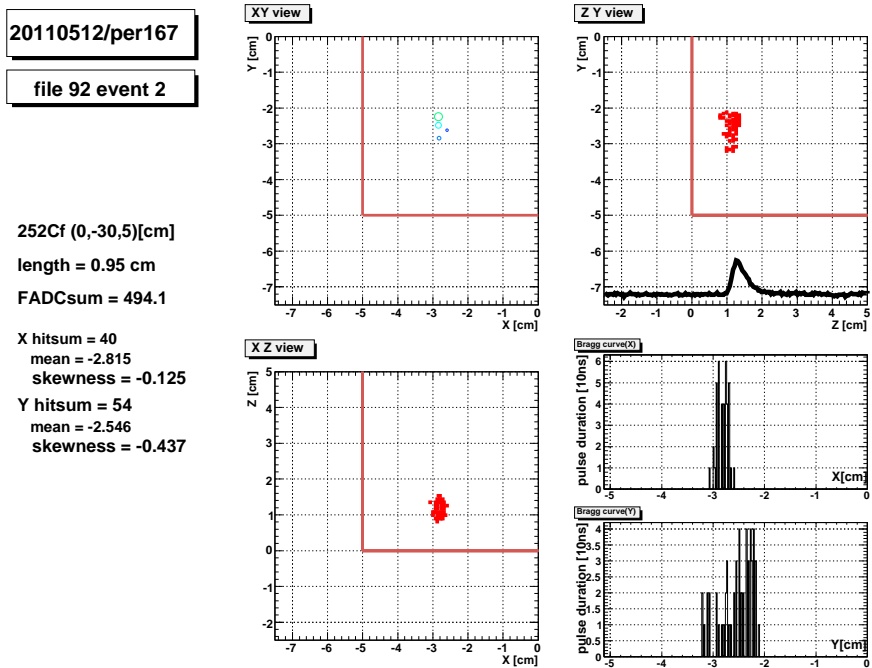}
\caption{\label{fig:track_92_2}
Typical event recorded with a improved DAQ system (TPC-mode5).
One set of raw data is a group of (X,Z) or (Y,Z) and is shown  with red square marks in the right-top and left-bottom panels. 
The pulse duration (time-over-threshold, TOT) at each strip is counted from the raw@data and we plot them as a function of each coordinate in the 
right-bottom panels. TOT have correlation with the charge 
With a software coincidence, X-Y track image is made, which is shown with open circles in the left-top panel. 
A size of each mark indicates the pulse duration of a coincidence hit.
The energy is 130keV.}
\end{figure}

We updated the FPGA firmware of our DAQ system in order to study the  
two-dimensional head-tail recognition with the NEWAGE-0.1a detector. 
We used to take X-Y coincidence in the FPGA at the rise of each hit.
We modified the firmware so that we do not require X-Y coincidence but 
record the rising and falling edges of all of the hit-strips(TPC-mode5).
We show a typical nuclear recoil event in Fig.~\ref{fig:track_92_2}. 
The energy is about 130 keV
\footnote{We calibrated the detector with alpha particle of about 1.5 MeV,
so we use an alpha-equivalent energy in this paper.}.

We defined a skewness ${\gamma}_{x_i}$ along ${x_i}=(x,y)$ axis by equation 
\ref{eq:skewness}, where $q(x_i)$ is the pulse duration at $x_i$ and $<x_i>$ is 
the mean value of $x_i$ with $q(x_i)>0$.

\begin{equation}
{\gamma}_{x_i}=\frac{<(q(x_i)\cdot(x_i-<x_i>)^3)>}{<(q(x_i)\cdot(x_i-<x_i>)^2)^{3/2}>}
\label{eq:skewness}
\end{equation}

\begin{figure}
\includegraphics[width=1.\linewidth]{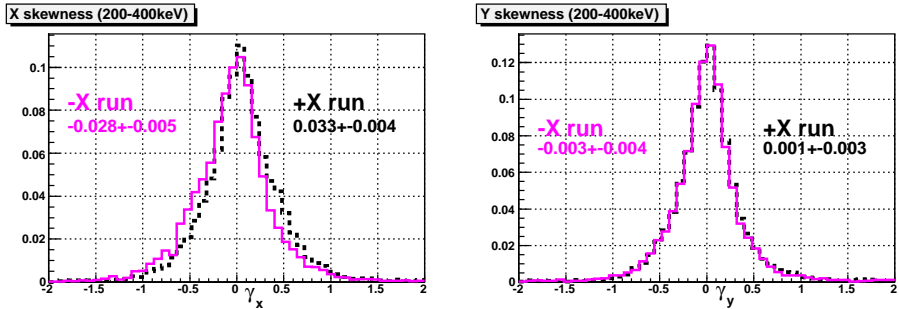}
\caption{\label{fig:skew_hist}Skewness distributions of +X run and 
-X run. The left and right panel shows the skewness along 
X axis (${\gamma}_{x}$) and Y(${\gamma}_{y}$) axis, respectively. 
The energy range is 200 - 400 keV. The center value of a 
Gaussian function and its fitting errors are indicated. 
The difference between the  
center values of +X  and -X runs 
are statistically significant,
while the center values of +Y run and -Y runs 
are statistically consistent.
}
\end{figure}

We measured the skewness by placing a 
$\rm {}^{252}Cf$ neutron source at four positions, 
(30cm, 0cm,5cm) (+X run), (-30cm, 0cm, 5cm) (-X run),  
(0cm, 30cm,5cm) (+Y run), and (0cm, -30cm,5cm) (-Y run).
(0cm, 0cm, 5cm) is the center of the detection volume and the 
$\mu$-PIC readout plane is the X-Y plane.
Skewness distributions of +X run and -X run are shown 
in Fig.~\ref{fig:skew_hist}
We fitted the distributions with a Gaussian function 
and we indicate the center value 
and its fitting errors in the figure. The difference between the  
center values of +X  and -X runs 
are statistically significant,
while the center values of +Y run and -Y runs 
are statistically consistent.


\begin{figure}
\includegraphics[width=.5\linewidth]{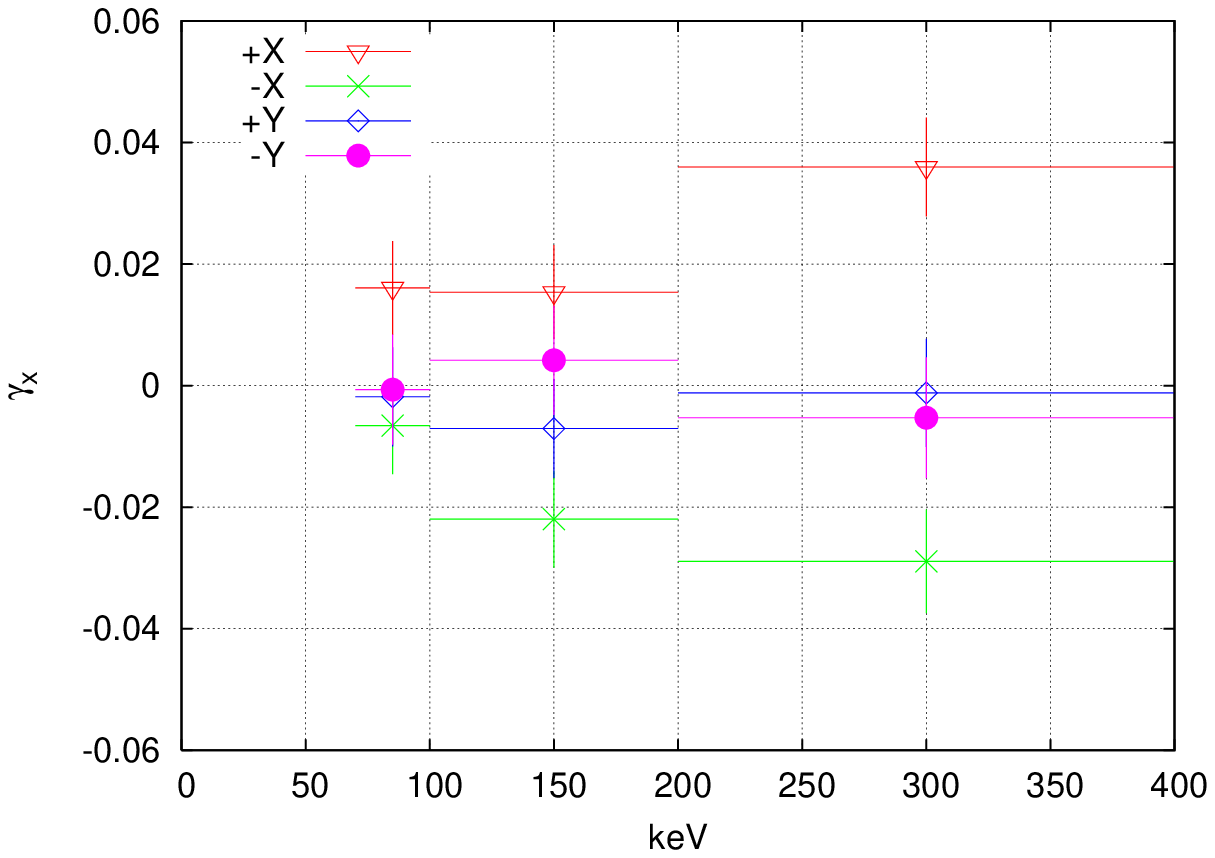}
\includegraphics[width=.5\linewidth]{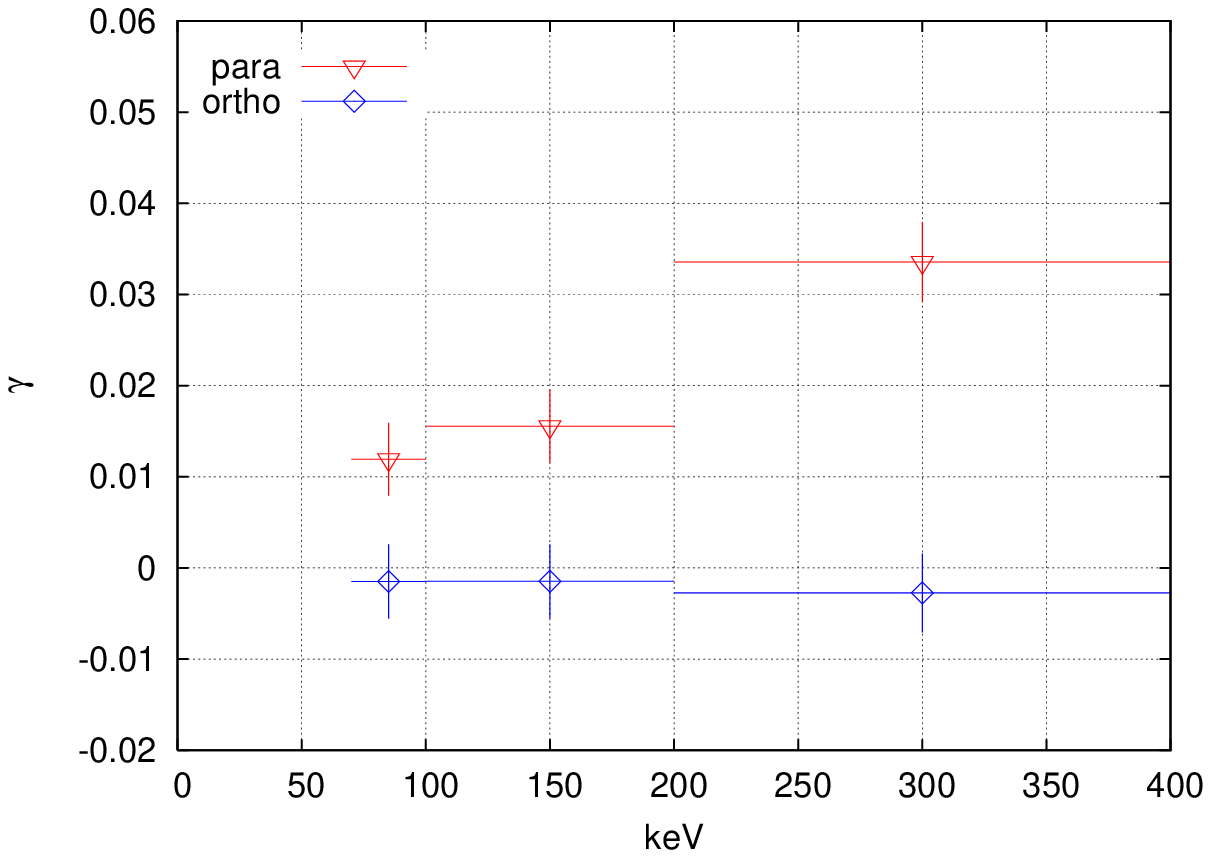}
\caption{\label{fig:e-dep}Energy dependences 
of ${\gamma}_x$ (left). 
Triangle (red online), cross (green), diamond (blue) and 
circle (violet) shows +X, -X, +Y, and -Y runs, respectively. 
The right panel shows the energy dependence of the combined 
skewness. The triangle (red online) and diamond (blue) show 
the skewness parallel and orthogonal to the incoming neutrons.
The number of events are  26586 (70-100keV), 21055 (100-200keV), and 
23283 (200-400keV). Only statistical errors are considered.}
\end{figure}

We show the energy dependence of ${\gamma}_x$ 
in the left panel Fig.~\ref{fig:e-dep}.
Though the statistics are not enough,  
positive (negative) skewness were observed 
in the +X (-X)  runs. 
We then combined these results into 
parallel( $\gamma_x$ in $\pm$X runs and $\gamma_y$ in $\pm$Y ) data and 
orthogonal( $\gamma_y$ in $\pm$X runs and $\gamma_x$ in $\pm$Y ) 
data.
In the parallel runs, absolute values of the skewness are used..
The result is shown in the right panel of Fig.~\ref{fig:e-dep}.
In the parallel runs, 
statically significant $\gamma$ with 3.0$\sigma$, 3.9$\sigma$, 
and  7.7 $\sigma$ are observed for the energy range of 
70-100keV, 100-200keV, and 200-400keV, respectively.
The  $\gamma$ were consistent with zero in the orthogonal runs.

Although the skewness definition is not optimized,  
these results shows that we can recognize head-tail 
with a sufficient statistics down to 70 keV.
These results also indicate much more efforts 
required for event-by-event recognition.

\section{R\&D for pixel-readout TPC}

TPCs for direction-sensitive dark matter search experiments need to 
detect short ($\rm \sim 1mm$) tracks with a reasonable ($\sim 30^o$)
angular resolution.
Most of these experiments use strip or MWPC 
readout,mostly because of the technical limitations. 
We are confronted with tracking difficulties
such as double-hit unfolding problems.
And thus none of us has achieved a required tracking performance with 
existing readout systems.
 A pixel-readout is an ideal solutions to these tracking difficulties
and would bring a breakthrough to this field.
Pixel-readout TPCs would realize the detection of 
the three-dimensional distribution of primary electrons, 
and thus we can in principle reconstruct the nuclear track 
without losing any information at the readout stage.
With these undeteriorated data,
a better angular resolution with a good direction uniformity is expected. 
A better head-tail recognition is also expected.
There are several achievements in the world to 
develop ASICs (application specified integrated circuits) 
for pixel readout TPCs.
TIMEPIX\cite{ref:TIMEPIX}, developed in CERN, is a well known PIXEL-readout 
ASIC which measures the time-of-flight (TOF) and the time-over-threshold (TOT)
with a pitch of 55 $\mu$m.
TOT is a good parameter to estimate collected charge as long as 
the longitudinal diffusion is small.
A dark matter detector needs to be as large as possible for a given detection 
area,  
and thus we want to have the drift region as long as possible. 
Then the longitudinal diffusion cannot be ignored and TOT no longer helps.
If we can have an ADC function in each pixel, 
though this is not very easy, 
this problem should be solved.
We can, furthermore use the ADC-TOT correlation to estimate 
the absolute z position.
Absolute z, even if the resolution is several cm, would greatly 
help to reduce the radioactive background from the drift plane, the GEM, 
and the $\mu$-PIC. 
 
We are developing a CMOS ASIC named QPIX which has ADC 
in addition to TOF and TOT in each pixel\cite{ref:Khoa_Mth}.
This development is in a starting phase and 
we are making efforts to prove the principle of concept and also evaluate the 
cost and background of the readout system.

\begin{figure}
\includegraphics[width=.5\linewidth]{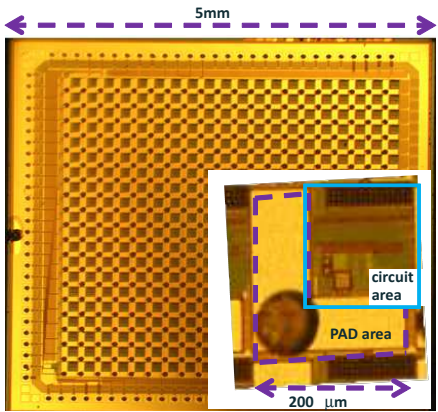}
\caption{\label{fig:QPIX-ver1}A microscope photo of QPIX-ver1. 20 $\times$ 20 
pixels are seen. The inset shows a zoom-up of one pixel.}
\end{figure}

After the R\&D with several types of TEGs (test element groups)
we developed QPIX-ver1, which is a first version with two-dimensional array.
The design values of QPIX-ver1 are shown in Table~\ref{tab:QPIX_spec} together with a goal values.
QPIX-ver1 has 20$\times$20 pixels with a pitch of 200 $\mu$m. Each pixel has 
14-bit TOF, 8-bit TOT, and 10 bit ADC. The chip was made by TSMC 0.18 
$\mu$m process. A microscope photo is shown in Fig. \ref{fig:QPIX-ver1}.  
20 $\times$ 20 pixels are seen. 
84 IO pads are placed along three edges of the chip.
The inset shows the zoom-up of one pixel. A metal pad area is 
indicated by dashed line. We can use this pad area for 
the direct charge collection from the gas volume 
or for the contact pad of bump bonding.
A trace of bump bonding test is seen in the center. The circuit area is 130 
$\times$ 130 $\rm\mu m^2$. 

\begin{table}
\begin{tabular}{|l|l|l|}
\hline
&QPIX-ver1 (design)& QPIX goal \\ \hline

dimension &	$\rm 200 \times 200{\mu}m^2$& $\rm 200 \times 200{\mu}m^2$\\
channels& 		20$\times$ 20ch/chip	   & \\
TOF& 	14 bits 	&14bits 	   \\
TOT& 	 8 bits&	8bits\\	   
comparator threshold(TOF,TOF) & 10fC &	1fC\\	   
clock(TOT,TOF) &100MHz &	100MHz\\
ADC& 	1.5pC / 10bits, 10Msps& 	100fC / 10bits, 10Msps\\	 
\hline
\end{tabular}
\caption{\label{tab:QPIX_spec}Design values of QPIX-ver1. The goal values are also shown.}
\end{table}

\begin{figure}
\includegraphics[width=1.\linewidth]{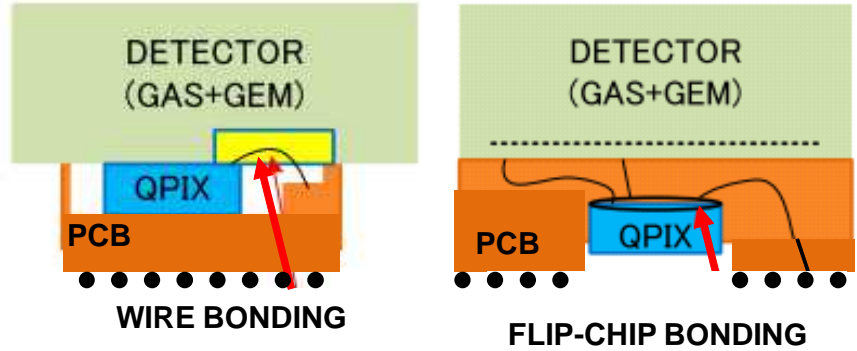}
\caption{\label{fig:QPIX_mountings}Two ways of QPIX mounting (side-views). 
The left panel shows a wire-bonding mounting and the right shows 
a flip-chip mounting.}
\end{figure}

We tried two ways of mountings, namely a wire-bonding mounting and 
a flip-chip mounting.
Both mounting ways are shown in Fig \ref{fig:QPIX_mountings}.
Wire-bonding methods is shown in the left panel.
This is a well-studied and very reliable method. We use wire-bondings 
to connect the IO pads of QPIX-ver1 to the readout PCB.
The problem of of the wire-bonding method is dead areas. 
We have dead areas at least along one edge with wire-bonding mountings. 

We tried another method, flip-chip mounting, in order to 
decrease the dead areas.
This method is shown in the right panel of Fig. \ref{fig:QPIX_mountings}.
We mount a charge collection PCB(CCPCB) on QPIX-ver1 by bump bonding.
The CCPCB has 20 $\times$ 20 pads on the gas side.
These pads are  connected to the cavity underneath through the CCPCB. 
QPIX-ver1 is mounted in this cavity by bump-bonding.
IO pad are also connected by bump-bonding and are connected to the  
mother broad PCB(MBPCB) through CCPCB.
The CCPCB is larger than QPIX-ver1 and can be mounted on a MBPCB
without dead areas.
We mounted four CCPCBs on a MBPCB. 
A mechanical mounting was confirmed, though electrical connection was not 
achieved. This was because 
the surface of the CCPCB cavity was not flat enough for the bump bonding.
We are trying to produce a better CCPCB to establish the flip-chip mounting.
We measured some performance of QPIX-ver1 mounted with the wire-bondings.

\begin{figure}
\includegraphics[width=.5\linewidth]{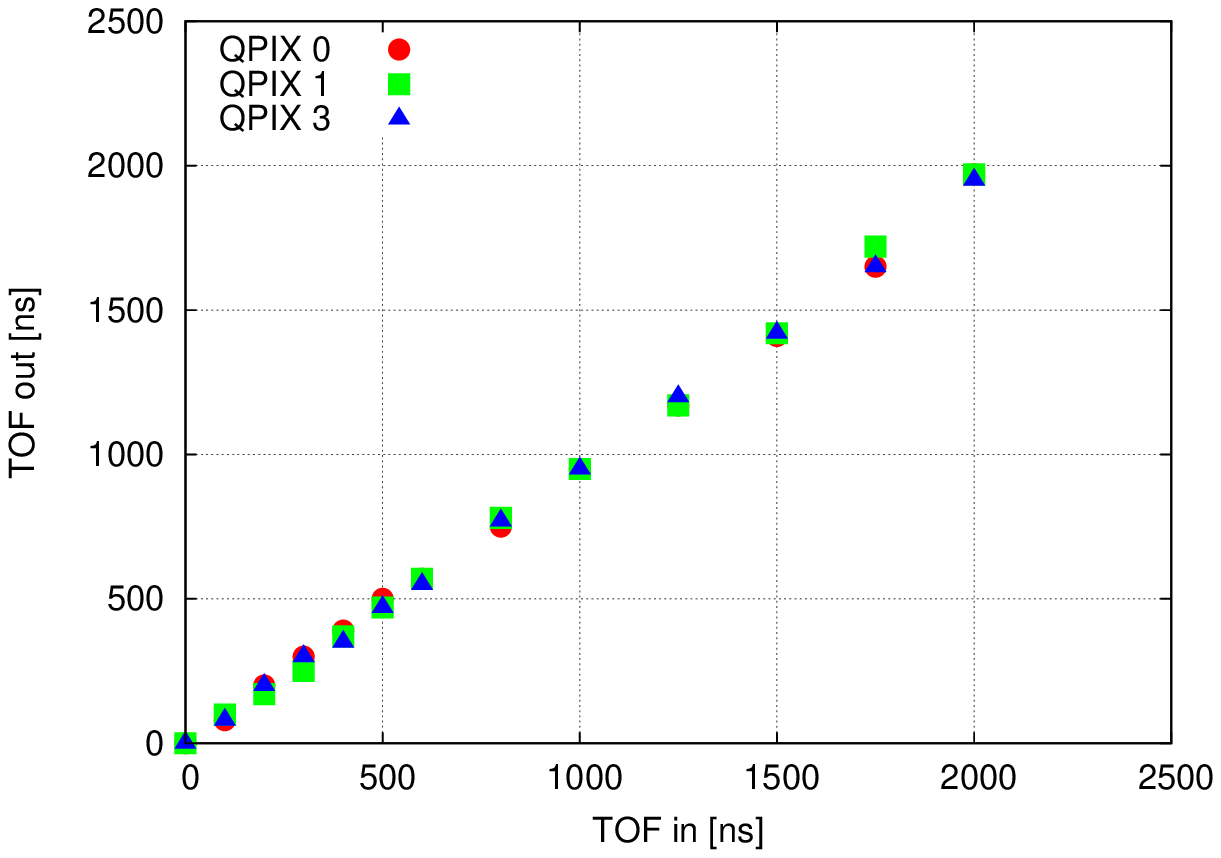}
\includegraphics[width=.5\linewidth]{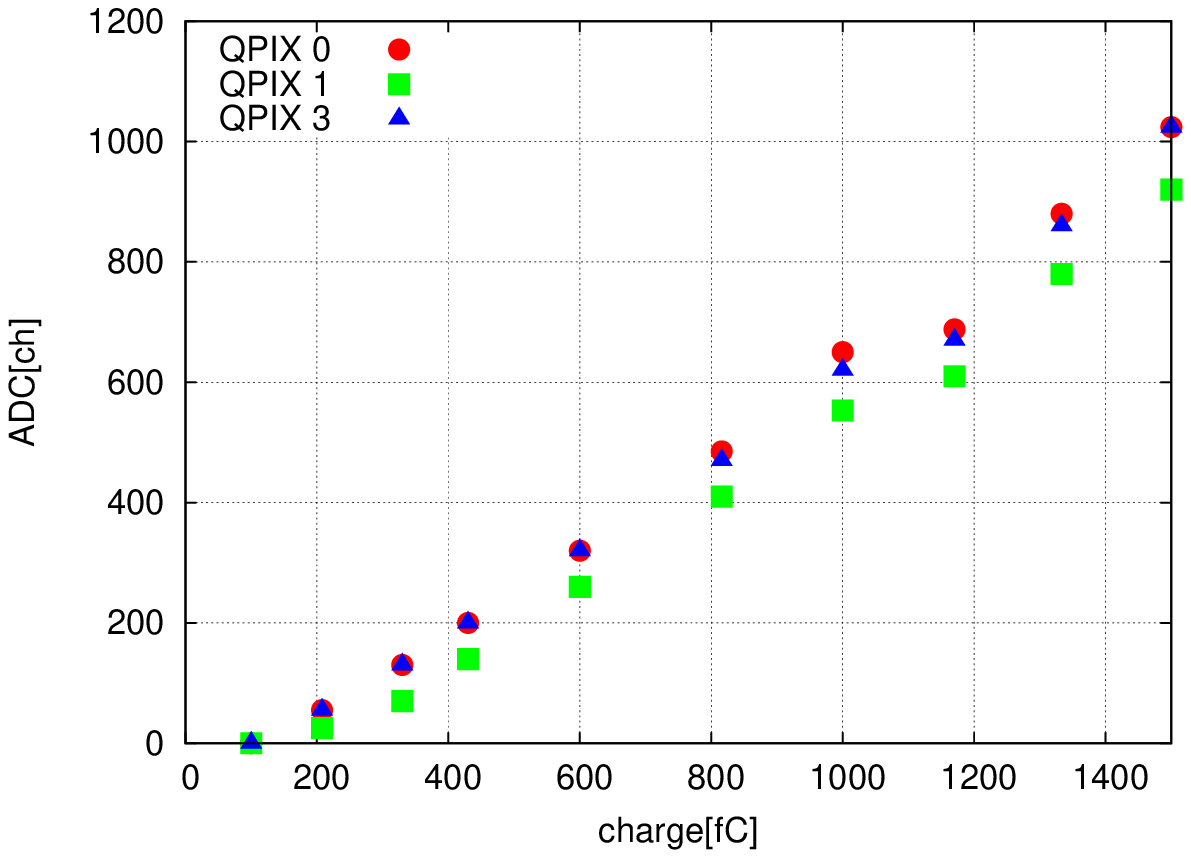}
\caption{\label{fig:QPIX_results}Four CCPCBs are mounted on a MBPCB. Inset shows a 
zoom-up of a CCPCB.}
\end{figure}

Measured performance of QPIX-ver1 is shown in Fig. \ref{fig:QPIX_results}.
Four QPIX chips are mounted.
Three chips worked, 
the rest had some trouble either in the ASIC development process or 
mounting process.
TOF shows good linearity up to 2$\mu$s.  
ADC shows fair linearity up to 1.5pC though the threshold was about ten times  
higher than the designed value. We are designing next TEG to improve the 
threshold. 

\section{Conclusions}
Intensive studies to improve the sensitivities of NEWAGE dark matter detectors
are under way.
We replaced the TPC cage with radio-pure PTFE.
We expect at least five times less count rate due to the radioactive 
background from the detector components.
We also studied the head-tail recognition in the surface laboratory 
using two-dimensional track data.
Although the skewness definition is not optimized,  
these results shows that we can recognize head-tail 
with a sufficient statistics down to 70 keV.
These results also indicate much more efforts 
required for event-by-event recognition.
For the future large volume detector, we are developing a pixel ASIC named
QPIX. We made a first version of arrayed pixels and tried two ways 
of mountings methods. 
We started a new dark matter run in August 2010 expecting a better 
limits. 

\section*{Acknowledgments}
This work was partially supported by KAKENHI (19684005, 23684014, 21340063, 
and 23654084).
\bibliographystyle{astron}

\bibliography{NEWAGE2011.bib}
\end{document}